\documentclass[10pt]{iopart} %%%%%%%%%%%%%%%%%%%%%%%%%%%%%%%%%%%%%%%%%%%%%%%%%%%%%%%%%%%%%%%%%%%%%%%%%%%%%%%%%%%%%%%%%%%%%%%%%%%%%%%%%%%%%%%%%%%%%%%%%%%%%%%%%%%%%%%%%%%%%%%%%%%%%%%%%%%%%%%%%%%%%%%%%%%%%%%%%%%%%%%%%%%%%%%%%%%%%%%%%%%%%%%%%%%%%%%%%%%%%%%%%%%%%%%%%%%%%%%%%%%%%%%%%%%% 
\usepackage{amsfonts} 
\usepackage{amssymb} 
\usepackage{graphicx}  

\begin{document}  
\title{The luminosity-redshift relation in brane-worlds:\\ I. Analytical results} 
\author{Zolt\'{a}n Keresztes, L\'{a}szl\'{o} \'{A}. Gergely, Botond Nagy, Gyula M. Szab\'{o} \\ 
EndAName Departments of Theoretical and Experimental Physics, University of Szeged, D\'om t\'er 9, Szeged 6720, Hungary}  

\begin{abstract} 
The luminosity distance - redshift relation is analytically given for generalized Randall-Sundrum type II brane-world models containing Weyl fluid either as dark radiation or as a radiation field from the brane. The derived expressions contain both elementary functions and elliptic integrals of the first and second kind. First we derive the relation for models with the Randall-Sundrum fine-tuning. Then we generalize the method for models with cosmological constant. The most interesting models contain small amounts of Weyl fluid, expected to be in good accordance with supernova data. The derived analytical results are suitable for testing brane-world models with Weyl fluid when future supernova data at higher redshifts will be available. 
\end{abstract}

\section{Introduction}  

At present the Universe is considered a general relativistic Friedmann space-time with flat spatial sections, containing more than $70\%$ dark energy and at about $25\%$ of dark matter. Dark energy could be simply a cosmological constant $\Lambda$, or quintessence or something entirely different. There is no widely accepted explanations for the nature of any of the dark matter or dark energy (even the existence of the cosmological constant remains unexplained).  

An alternative to introducing dark matter would be to modify the law of gravitation, like in MOND \cite{MOND} and its relativistic generalization  \cite{MONDrel}. These theories are compatible with the Large scale structure of the Universe \cite{MOND2}. However in spite of the successes, certain problems were signaled on smaller scales \cite{MOND3}.  

Quite remarkably, supernova data, which in the traditional interpretation yield to the existence of dark energy, can be explained by certain f(R) \cite{fR} or inverse curvature gravity models \cite {modgrav}. However the parameter range, in which the latter is in goood agrement with the supernova data, also presents stability problems \cite{excmodgrav}. 

Modifications of the gravitational interaction could also occur by enriching the space-time with extra dimensions. Originally pioneered by Kaluza and Klein, such theories contained compact extra dimensions. The so-called brane-world models, motivated by string / M-theory, containing our observable 4-dimensional universe (the brane) as a hypersurface, were introduced in \cite{ADD}, \cite{RS1} and \cite{RS2}, the latter model allowing for a non-compact extra dimension.  

The curved generalizations of the model presented in \cite{RS2} have evolved into a 5-dimensional alternative to general relativity, in which gravity has more degrees of freedom. In contrast with standard model fields, these evolve in the whole 5-dimensional bulk. In this generalized Randall-Sundrum type II (RS) theory, the brane has a tension $\lambda $ and gravitational dynamics is governed by the 5-dimensional Einstein equation. Its projections to our observable 4-dimensional universe (the brane) are the twice contracted Gauss equation, the Codazzi equation and an effective Einstein equation, the latter being obtained by employing the junction conditions across the brane \cite{Decomp}. The effective Einstein equation (for the case of symmetric embedding and no other contribution to the bulk-energy-momentum than a bulk cosmological constant) was first given in a covariant form in \cite{SMS}. Supplementing this by the pull-back to the brane of the bulk energy momentum tensor $\widetilde{\Pi }_{ab}$, which is  
\begin{equation} \mathcal{P}_{ab}=\frac{2\widetilde{\kappa }^{2}}{3}\left( g_{a}^{c}g_{b}^{d} \widetilde{\Pi }_{cd}\right) ^{TF} 
\end{equation} 
(with $\widetilde{\kappa }^{2}$ the bulk coupling constant and $g_{ab}$ the induced metric on the brane) the effective Einstein equation reads \cite{Decomp}:  
\begin{equation} G_{ab}=-\Lambda g_{ab}+\kappa ^{2}T_{ab}+\widetilde{\kappa }^{4}S_{ab}- \mathcal{E}_{ab}+\mathcal{P}_{ab}\ .  \label{modE} \end{equation}
Here $\kappa ^{2}$ is the brane coupling constant, related to the bulk coupling constant and the brane tension $\lambda $ as $6\kappa ^{2}=\widetilde{\kappa }^{4}\lambda $, and  
\begin{equation} \Lambda =\frac{\kappa^{2}}{2} \lambda -\frac{\widetilde{\kappa }^{2}}{2} n^{c}n^{d} \widetilde{\Pi }_{cd} 
\end{equation} 
represents a cosmological "constant" which possibly varies due to the normal projection of the bulk energy-momentum tensor (this includes the contribution $-\widetilde{\Lambda }g_{ab}$ due to the bulk cosmological constant $\widetilde{\Lambda }$). The source term $S_{ab}$ is quadratic in the brane energy-momentum tensor $T_{ab}$:  
\begin{equation} S_{ab}=\frac{1}{4}\Biggl[-T_{ac}^{\ }T_{b}^{c}+\frac{1}{3}TT_{ab}-\frac{ g_{ab}}{2}\left( -T_{cd}^{\ }T^{cd}+\frac{1}{3}T^{2}\right) \Biggr]\ , \label{S} 
\end{equation} 
and $\mathcal{E}_{ab}$ is the electric part of the bulk Weyl tensor $ \widetilde{C}_{abcd}$, given as  \begin{equation} \mathcal{E}_{ac}=\widetilde{C}_{abcd}n^{b}n^{d}\ .  \label{calE} 
\end{equation} 
In a cosmological context and suppressing any energy exchange between the brane and the bulk, this latter term generates the so-called dark radiation. Otherwise it can be called a Weyl fluid.  

A review of many aspects related to the theories described by the effective Einstein equation (\ref{modE}) can be found in \cite{MaartensLR}. Both early cosmology \cite{BDEL} and gravitational collapse \cite{BGM}-\cite{Pal} are essentially modified in these theories. There is also possible to replace dark matter with geometric effects in the interpretation of galactic rotation curves, weak lensing and galaxy cluster dynamics \cite{Harko}.  

\begin{figure}[tbp] 
\includegraphics[height=8cm]{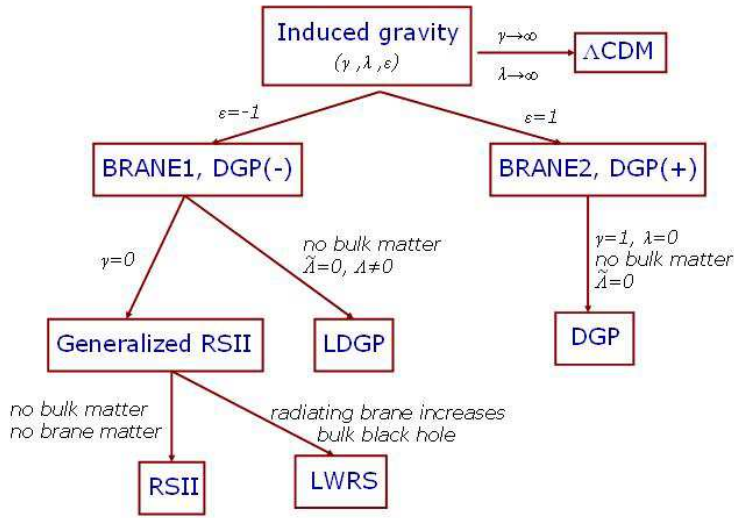} 
\caption{(Color online) A diagram presenting various brane-world models and their inter-relations. LWRS is the generalized Randall-Sundrum model with cosmological constant and a Weyl fluid reflecting a brane radiating into the bulk during nowadays or at least until recent cosmological times.} 
\label{Fig1} 
\end{figure}

The possible modifications of gravitational dynamics are even more versatile in the so-called induced gravity models. These can be regarded as brane-world models enhanced with the first quantum-correction arising from the interaction of the brane matter with bulk gravity. The induced gravity correction couples to the 5-dimensional Einstein-Hilbert action with the coupling constant $\gamma \widetilde{\kappa }^{2}/\kappa ^{2}$. The simplest of such models, the DGP model was introduced in \cite{DGP}. This model however suffers from linear instabilities (ghost modes in the perturbations), as shown for de Sitter branes \cite{ghost}. The ghost modes withstand even the introduction of a second brane \cite{ghost2}. Generalizations of the DGP model are discussed covariantly in \cite{SS} and  \cite{MMT} when the embedding is symmetric, and in \cite{Induced} when it is asymmetric. In these models the role of the effective Einstein equation (\ref{modE}) is taken by a more complicated equation (see for example Eq. (29) of  \cite{Induced}), which contains the square of the Einstein tensor $G_{ab}$. This implies that in certain sense the degree of nonlinearity of the theory is squared. In a cosmological setup the square root of this equation can be taken, leading to a set of modified Friedmann and Raychaudhuri equations, which however contain a sign ambiguity $\varepsilon =\pm 1$ due to the involved square root. These are called the BRANE1 [DGP(-)] branch for $\varepsilon =-1$ and BRANE2 [DGP(+)] for $\varepsilon =1$ in the terminology of \cite{SS} [or \cite{LMM}, respectively]. Both the original Randall-Sundrum type II model and the DGP model are contained as special subcases. Notably, the BRANE2 branch contains cosmological models which self-accelerate at late-times. We give in Fig \ref{Fig1} a diagram containing a classification of these theories and how they emerge as different limits from each other.  

In this paper we discuss analytically the luminosity distance - redshift relation in various generalized Randall-Sundrum type II brane-world models described by Eq. (\ref{modE}). Our analytical approach can enhance the confrontation of these models with current and most notably, with future supernova observations. We note that recently analytical results have been given in Ref. \cite{DabrowskiS} for a wide class of phantom Friedmann cosmologies too, in terms of elementary and Weierstrass elliptic functions.  

In section 2 we review the notion of luminosity distance, its relation with the redshift and how these can be measured independently. This section was included mainly for didactical purposes.  

In section 3 we review the modification of this relation in the Randall-Sundrum type II brane-world scenario. These include the introduction of the parameters $\Omega _{\lambda }$ and $\Omega _{d}$ which can be traced back to the source terms $S_{ab}$ and $\mathcal{E}_{ab}$ of the modified Einstein equation (\ref{modE}). The other cosmological parameters are $\Omega _{\rho }$, representing (baryonic and dark) matter and $\Omega _{\Lambda }$. We do not include bulk sources in the analysis, with the notable exception of a bulk cosmological constant.  

Section 4 contains the derivation of the analytic expression for the luminosity distance - redshift relation for the brane-worlds which are closest to the original Randall-Sundrum scenario \cite{RS2}, thus with no cosmological constant (Randall-Sundrum fine-tuning). The generic expression (\ref{solution1}) of the luminosity distance derived here is given in terms of elementary functions and elliptic integrals of the first and second kind. From this most generic case we take the subsequent limits: $\Omega _{d}=0$ (subsection 4.2), $\Omega _{\lambda }=0$ (subsection 4.3); and both $\Omega _{d}=\Omega _{\lambda }=0$, this being the general relativistic Einstein-de Sitter case (subsection 4.4).  

Such models however could not allow for late-time acceleration, therefore in section 5 we discuss the luminosity distance - redshift relation for brane-worlds with $\Lambda $. First we present in subsection 5.1 a class of models, for which the luminosity distance can be given in terms of elementary functions alone. These models are characterized by an extremely low value of the brane tension, thus are in conflict with various constraints on brane-world models.  

Next, in subsection 5.2 we discuss brane-worlds for which the brane-characteristic contributions $\Omega _{\lambda }$ and $\Omega _{d}$ represent small perturbations. This is a good assumption as observational evidences suggest that general relativity is a sufficiently accurate theory of the universe, and as such the deviations from it could not be very high, at least at late-times. We give analytical expressions in terms of both elementary functions and elliptic integrals of the first and second kind for the luminosity distance, to first order accuracy in the chosen small parameters of the model. Some of the most lengthy computations needed in order to achieve the result are presented in the Appendix.  

Section 6 contains the concluding remarks.  

Throughout the paper $c=1$ was employed.

\section{The luminosity-redshift relation}  

The Friedmann-Lema\^{\i}tre-Robertson-Walker (FLRW) metric 
\begin{equation} 
ds_{FLRW}^{2}=-d\tau ^{2}+a^{2}\left( \tau \right) \left[ \frac{dr^{2}}{ 1-kr^{2}}+r^{2}\left( d\theta ^{2}+\sin ^{2}\theta d\varphi ^{2}\right)  \right] \   \label{FLRW} 
\end{equation}
describes a homogeneous and isotropic universe. Here $\tau $ is cosmological time, ($r,\,\theta ,\,\varphi $)$\ $\ are comoving coordinates, $a$ is the scale factor and $k=0,\,\pm 1$ the curvature index. The \textit{proper radial distance} is defined as $ar$. A useful alternative form of the FLRW metric is 
\begin{equation} 
ds_{FLRW}^{2}=-d\tau ^{2}+a^{2}\left( \tau \right) \left[ d\chi ^{2}+ r^{2}\left( \chi ;k\right) \left( d\theta ^{2}+\sin ^{2}\theta d\varphi ^{2}\right) \right] \ ,  \label{FLRW1} \end{equation} 
with 
\begin{equation} 
r=r(\chi ;k)=\left\{  \begin{array}{cc} \sin \ \chi \  & ,\qquad k=1\,, \\  \chi & ,\qquad k=0\,, \\  \sinh \ \chi & \ ,\qquad k=-1\,. \end{array} \right. \,, 
\end{equation} 
$\chi $ being an other comoving radial coordinate.  

\begin{figure}[tbp] 
\includegraphics[height=8cm]{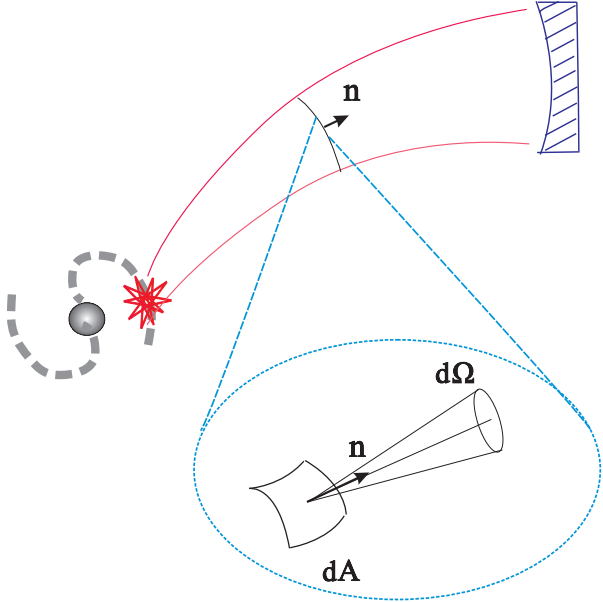} 
\caption{(Color online) A schematic representation of the propagation in the curved space-time of the light emitted by a supernova explosion in a distant galaxy and collected on the telescope mirror. A \textit{dimensional magnification} (\`{a} la Wheeler) shows the elementary area $dA$ normal to the direction of propagation $\mathbf{n}$ and the elementary solid angle $d\Omega $ around $\mathbf{n}$. } 
\label{Fig2} 
\end{figure}

If a photon stream emitted by an astrophysical light source travel without collisions, the number of photons $dN_{\gamma }$ from a comoving elementary volume of the $6$-dimensional phase space ($\vec{x},\ \vec{p}$) is conserved  \cite{Padmanabhan}. Thus the phase space density  
\begin{equation} 
f(t,\vec{x},\vec{p})\equiv \frac{dN}{d^{3}\vec{x}d^{3}\vec{p}}=\frac{dN}{ \omega ^{2}d\tau \ dA\ d\omega \ d\Omega }\   \label{f} 
\end{equation} 
of a photon stream is constant in time. Here $\omega $ denotes the frequency of the photons, $dA$ and $d\Omega $ stand for the elementary area normal to the direction of propagation and for the elementary solid angle around the direction of propagation, respectively (see Fig \ref{Fig2}). Eq. (\ref{f}) holds true for any kind of cosmological evolution, provided $d^{3}\vec{x} \propto d\tau dA$ and $d^{3}\vec{p}\propto \omega ^{2}d\omega d\Omega $ are valid for the photons \cite{Padmanabhan}. The \textit{luminosity} of the source is $\mathcal{L}=dE_{em}/dt_{em}$ (total energy produced in unit time; the suffix $em$ refers to emission).

A telescope detects the \textit{photon flux} $\mathcal{F}=dE_{rec}/d\tau _{rec}/A_{M}$ (the suffix $rec$ refers to reception). This is the energy detected during unit time on the telescope mirror surface $A_{M}$. (The surface $A_{M}$ is understood to be perpendicular to the incident light stream.)  

From their definition, one can easily find a relation between $\mathcal{F}$ and $\mathcal{L}$:  
\begin{equation} 
\frac{\mathcal{F}A_{M}}{\mathcal{L}}=\frac{dE_{rec}/d\tau _{rec}}{dE_{em}/d\tau _{em}}\,. 
\end{equation} 
As the energy of the photon stream in the comoving elementary phase space volume is $dE=\hbar \omega \ dN$, from Eq. (\ref{f}) we find  \begin{equation} 
\frac{dE_{rec}}{dE_{em}}=\frac{A_{M}}{A_{tot}}\left( \frac{\omega _{rec}}{\omega _{em}}\right) ^{3}\frac{d\omega _{rec}}{d\omega _{em}}\frac{dA_{rec}}{ dA_{em}}\frac{d\tau _{rec}}{d\tau _{em}}\,.  \label{dEdE} 
\end{equation} 
Here we have used that from the isotropy of the FLRW universe $d\Omega _{rec}=d\Omega _{em}$ and\ we integrate the first to the solid angle encompassing the mirror surface, the second to the whole solid angle (cf. the definitions of $E_{rec}$, $E_{em}$). In Eq. (\ref{dEdE}) $A_{tot}$ represents the \textit{proper} area of a sphere centered in the light source and containing the reception point on its surface, at the time of reception.  

Due to cosmological evolution the elementary area $dA$ changes as $a^{2}$ and the frequency of the light is redshifted during cosmic expansion, $\omega \,\propto \,1/a$ \cite{Padmanabhan}. In the cosmological evolution of the comoving elementary phase space volume element $d\omega $ changes accordingly: $d\omega \,\propto \,1/a$. Therefore  
\begin{equation} 
\frac{\mathcal{F}}{\mathcal{L}}=\frac{1}{A_{tot}}\left( \frac{a}{a_{0}} \right) ^{2}\,, 
\end{equation} 
where $a_{0}$ is the present value of the scale factor, and$\ a$ is understood to be the scale factor at emission time. In the FLRW universe the proper area of a sphere with comoving radius $r_{em}$ is $A_{tot}=4\pi a_{0}^{2}r_{em}^{2}$, and the redshift $z$ is defined as \begin{equation} 
1+z=\frac{a_{0}}{a_{em}}\ .  \label{z} 
\end{equation} 
The \textit{luminosity distance} $d_{L}$ is defined as in Euclidean geometry:  
\begin{equation} 
d_{L}(z):=\left( \frac{\mathcal{L}}{4\pi \mathcal{F}}\right) ^{1/2}=a_{0}r_{em}(1+z)\ .  \label{dl1} 
\end{equation} 
This definition is rigorous as long as we are dealing with the (homogeneous and isotropic) FLRW universe (irrespective of the value of the curvature index $k$) and the radius of a sphere is measured in the proper distance $ra$ (the FLRW metric (\ref{FLRW}) guarantees that the surface of a sphere with radius $ra$ is $4\pi a^{2}r^{2}$).  

According to Eq. (\ref{FLRW1}) the comoving coordinate $r_{em}$ can be written in terms of an other radial comoving coordinate $\chi _{em}$ (representing the location of the source): 
\begin{equation} d_{L}(z)\equiv a_{0}\left( 1+z\right) \mathcal{H}\left( \chi _{em};k\right) \ .  \label{lumred} 
\end{equation} 
Disregarding possible deflections by perturbations of the FLRW universe, a light ray follows radial null geodesics of the FLRW metric, characterized by  $d\chi =d\tau /a(\tau )=da/a^{2}H.$ Here $H=\dot{a}/a$ is the Hubble parameter. Then 
\begin{equation} 
\chi _{em}=\chi \left( a_{em}\right) =\int_{a_{em}}^{a_{0}}\frac{da}{a\dot{a} }=\int_{a_{em}}^{a_{0}}\frac{da}{a^{2}H\left( a\right) }\mathrm{{\ }.} \label{chi} 
\end{equation} 
By employing Eq. (\ref{z}) the radial variable $\chi $ can also be expressed in terms of an integral over the redshift as 
\begin{equation} \chi _{em}\left( z\right) =\frac{1}{a_{0}}\int_{0}^{z}\frac{dz^{\prime }}{ H\left( z^{\prime }\right) }\ ,  \label{chiz} 
\end{equation} 
which completes the definition (\ref{lumred}) of the luminosity distance $d_{L}$ in terms of the redshift $z$.  

Differentiating Eq. (\ref{lumred}) with $\chi $ given by Eq. (\ref{chiz}) with respect to $z$ gives 
\begin{equation} 
\frac{1}{H(z)}=\left[ 1-\frac{kd_{L}^{2}(z)}{a_{0}^{2}(1+z)^{2}}\right] ^{-1/2}\ \frac{d}{dz}\left[ \frac{d_{L}(z)}{1+z}\right] \ , 
\end{equation} 
therefore if independent measurements of $d_{L}$ and $z$ are available for a set of light sources, the Hubble-parameter $H(z)$ and in consequence the cosmological dynamics can be determined.  

From the combined measurements of the large-scale structure of the Universe  \cite{SDSS k=0}, \cite{SDSS eisenstein} and of the structure of the cosmic microwave background \cite{SDSS WMAP k=0} the conclusion was reached that the space geometry has flat spatial sections. Therefore in what follows we consider $k=0$. Then the luminosity distance-redshift relation becomes 
\begin{equation} d_{L}(z)=\left( 1+z\right) \int_{0}^{z}{\frac{dz^{\prime }}{H(z^{\prime })}~.}  \label{dl2} 
\end{equation}  

In practice, the function $d_{L}(z)$ is conveniently measured with distant supernovae of type Ia. The luminosity is evaluated by photometry, while the redshift from spectroscopic analysis of the host galaxy.  

Each cosmological model has its own prediction for the shape of the function  $d_{L}(z)$ [see Eq. (\ref{lumred}) with $\chi $ given by Eq. (\ref{chiz}) for generic $k$, or Eq. (\ref{dl2}) for $k=0$]. This is how the measured $d_{L}(z)$ data turn into a cosmological test.

\section{The luminosity-redshift relation in Randall-Sundrum type II brane-worlds}  

We consider FLRW branes with $k=0$ and brane cosmological constant $\Lambda $, embedded symmetrically. The bulk is the Vaidya-anti de Sitter space-time with cosmological constant $\widetilde{\Lambda }$, and it contains bulk black holes with masses $m$ on both sides of the brane. The black hole masses can change if the brane radiates into the bulk. An ansatz comparable with structure formation has been advanced in \cite{PalStruct} for the Weyl fluid $m/a^{4}$ for the case when the brane radiates, $m=m_{0}a^{\alpha }$, where $m_{0}$ is a constant and $\alpha =2,~3$. For $\alpha =0$ the Weyl fluid is known as dark radiation and then the bulk space-time becomes Schwarzschild-anti de Sitter. The brane tension and the two cosmological constants are inter-related as  
\begin{equation} 
2\Lambda =\kappa ^{2}\lambda +\widetilde{\kappa }^{2}\widetilde{\Lambda }~. \label{finetune} 
\end{equation}  

The Friedmann equation gives the Hubble parameter to $\Lambda $, $m$,$\ $the scale factor $a$ and the matter energy density $\rho $ on the brane:  
\begin{equation} 
H^{2}=\frac{\Lambda }{3}+\frac{\kappa ^{2}\rho }{3}\left( 1+\frac{\rho }{2\lambda }\right) +\frac{2m_{0}}{a^{4-\alpha }}\mathrm{{\ }.} \label{Friedmann} 
\end{equation} 
In the matter dominated era the brane is dominated by dust, obeying the continuity equation 
\begin{equation} 
\dot{\rho}+3H\rho =0\mathrm{{\ },} 
\end{equation} 
which gives $\rho \sim a^{-3}$. We introduce the following dimensionless quantities:  
\begin{eqnarray} 
\Omega _{tot} &=&\Omega _{\Lambda }+\Omega _{\rho }+\Omega _{\lambda }+\Omega _{d}~,  \label{sumom} \\ \Omega _{\rho } &=&\frac{\kappa ^{2}\rho _{0}}{3H_{0}^{2}}\ ,\ \ \ \ \ \Omega _{\lambda }=\frac{\kappa ^{2}\rho _{0}^{2}}{6\lambda H_{0}^{2}}\ , \\ \Omega _{d} &=&\frac{2m_{0}}{a_{0}^{4-\alpha }H_{0}^{2}}\mathrm{\ },\ \ \ \ \ \Omega _{\Lambda }=\frac{\Lambda }{3H_{0}^{2}}\ .  \label{omd} 
\end{eqnarray} 
The subscript $0$ denotes the present value of the respective quantities. In terms of these notations the Friedmann equation becomes  \begin{equation} 
H^{2}=H_{0}^{2}\left[ \Omega _{\Lambda }+\Omega _{\rho }\frac{a_{0}^{3}}{ a^{3}}+\Omega _{d}\frac{a_{0}^{4-\alpha }}{a^{4-\alpha }}+\Omega _{\lambda } \frac{a_{0}^{6}}{a^{6}}\right] \ . 
\end{equation} 
In particular at present time this gives $\Omega _{tot}=1$. Then the radial coordinate (\ref{chi}) becomes  
\begin{equation} 
\chi _{em}=\frac{1}{H_{0}}\int_{a_{em}}^{a_{0}}\frac{ada}{\left[ \Omega _{\Lambda }a^{6}+\Omega _{\rho }a_{0}^{3}a^{3}+\Omega _{d}a_{0}^{4-\alpha }a^{\alpha +2}+\Omega _{\lambda }a_{0}^{6}\right] ^{1/2}}\ .  \label{chi2} 
\end{equation} 
This is a complicated integral, which cannot be computed analytically in the majority of cases. In what follows we will analyze various specific cases of the above integral, when an analytic solution is possible. The cases $\alpha =2,~3$ represent the Weyl fluid compatible with structure formation, while $\alpha =0$ represents the dark radiation.

\section{Branes with Randall-Sundrum fine-tuning}  

In the original Randall-Sundrum scenario the bulk cosmological constant $ \widetilde{\Lambda }$ is fine-tuned with the brane tension $\lambda $ such that cf. Eq. (\ref{finetune}) the brane cosmological constant vanishes. For simplicity we also assume throughout this section $\alpha=0$.
By imposing a vanishing cosmological constant on the brane, $\Omega _{\Lambda }=0$ such that the polynomial of rank $6$ in the denominator of the integrand in Eq. (\ref{chi2}) shrinks to a polynomial of rank $3$. Therefore its roots can be found analytically. Following general procedures, the luminosity distance - redshift relation can be then given analytically in terms of elliptic functions. This is done in the following subsection. In the second and third subsections of this chapter we discuss the limits $\Omega _{d}\rightarrow 0$ (when the bulk is anti de Sitter) and the late-time universe limit $\rho /\lambda \rightarrow 0 $. The general relativistic (Einstein-deSitter) limit is found in the fourth subsection, when further $\Omega _{\lambda }\rightarrow 0$ is taken.

\subsection{Schwarzschild-AdS bulk}  

With no brane cosmological constant, Eq. (\ref{chi2}) becomes: 
\begin{equation} 
\chi _{em}=\frac{1}{a_{0}^{3/2}H_{0}\Omega _{\rho }^{1/2}} \int_{a_{em}}^{a_{0}}\frac{ada}{\left[ a^{3}+\frac{\Omega _{d}}{\Omega _{\rho }}a_{0}a^{2}+\frac{\Omega _{\lambda }}{\Omega _{\rho }}a_{0}^{3} \right] ^{1/2}}\mathrm{.}  \label{chi4} 
\end{equation} 
Following the method given in \cite{Bronstein} we find the following roots of the denominator: 
\begin{eqnarray} 
\alpha &=&-\frac{2\Omega _{d}a_{0}}{3\Omega _{\rho }}\left( 1+2\cosh \frac{\Psi }{3}\right) \mathrm{,}  \nonumber \\ 
\beta &=&\frac{\Omega _{d}a_{0}}{3\Omega _{\rho }}\left( -1+\cosh \frac{\Psi  }{3}+i\sqrt{3}\sinh \frac{\Psi }{3}\right) \mathrm{,} \label{roots} 
\end{eqnarray}
and its complex conjugate $\beta ^{\ast }$. 
The auxiliary quantity $\Psi $ is defined as 
\begin{equation} 
\cosh \Psi =1+\frac{27\Omega _{\lambda }\Omega _{\rho }^{2}}{2\Omega _{d}^{3}}\mathrm{.}  \label{Psi} 
\end{equation} 
We introduce the following \textit{real} combinations of the complex roots 
\begin{eqnarray} 
b_{1} &=&\frac{\beta +\beta ^{\ast }}{2}=\frac{\Omega _{d}a_{0}}{3\Omega _{\rho }}\left( -1+\cosh \frac{\Psi }{3}\right) \mathrm{,}  \nonumber \\ 
a_{1} &=&\frac{\beta -\beta ^{\ast }}{2i}=\frac{\Omega _{d}a_{0}}{\sqrt{3}\Omega _{\rho }}\sinh\frac{\Psi }{3} \mathrm{.}  \label{a1b1} 
\end{eqnarray} 
Then Eq. (\ref{chi4}) is written conveniently as 
\begin{equation} 
\chi _{em}=\frac{1}{a_{0}^{3/2}H_{0}\Omega _{\rho }^{1/2}} \int_{a_{em}}^{a_{0}}\frac{ada}{\left( a-\alpha \right) \left[ \left( a-b_{1}\right) ^{2}+a_{1}^{2}\right] ^{1/2}}\mathrm{.}  \label{chiroots} 
\end{equation} 
The integration can be carried out by employing the formulae (239.07) and (341.53) of Ref. \cite{Byrd}. We obtain 
\begin{eqnarray} 
\chi _{em} &=&\frac{1}{a_{0}^{3/2}H_{0}\Omega _{\rho }^{1/2}}\Biggl\{g\left( \alpha +A\right) \left[ F\left( \varphi _{0},\varepsilon \right) -F\left( \varphi _{em},\varepsilon \right) \right]  \nonumber \\ &&-2gA\left[ E\left( \varphi _{0},\varepsilon \right) -E\left( \varphi _{em},\varepsilon \right) \right]  \nonumber \\ &&+2gA\Biggl[\frac{\sin \varphi _{0}\sqrt{1-\varepsilon ^{2}\sin ^{2}\varphi _{0}}}{1+\cos \varphi _{0}}  \nonumber \\ &&-\frac{\sin \varphi _{em}\sqrt{1-\varepsilon ^{2}\sin ^{2}\varphi _{em}}}{ 1+\cos \varphi _{em}}\Biggr]\Biggr\}\mathrm{,} 
\end{eqnarray} 
where $F\left( \varphi ,\varepsilon \right) $ is the elliptic integral of the first kind; $E\left( \varphi ,\varepsilon \right) $ is the elliptic integral of the second kind (with variable $\varphi $ and argument $\varepsilon $); and we have introduced the following standard notations, cf. Ref. \cite{Byrd}:  
\begin{eqnarray} 
A^{2} &=&\left( b_{1}-\alpha \right) ^{2}+a_{1}^{2}\mathrm{,}  \nonumber \\ \varepsilon ^{2} &=&\frac{A+b_{1}-\alpha }{2A}\mathrm{,}  \nonumber \\ g &=&\frac{1}{\sqrt{A}}\mathrm{,}  \nonumber \\ \varphi &=&\arccos \left( \frac{A+\alpha -a}{A-\alpha +a}\right) \mathrm{.} 
\end{eqnarray} By employing Eqs. (\ref{z}), (\ref{chiz}) and (\ref{dl2}), after a lengthy, but straightforward calculation, the luminosity distance-redshift relation emerges: 
\begin{eqnarray} 
d_{L}^{\lambda d} &=&\frac{\left( 1+z\right) \Omega _{d}^{1/2}}{H_{0}\Omega _{\rho }}\Biggl\{\left( B_{1}+B_{2}\right) \left[ F\left( \varphi _{0},\varepsilon \right) -F\left( \varphi _{em},\varepsilon \right) \right]  \nonumber \\ &&+2B_{2}{}{}{}\Biggl[E\left( \varphi _{em},\varepsilon \right) -E\left( \varphi _{0},\varepsilon \right)  \nonumber \\ &&+\frac{\sin \varphi _{0}\sqrt{1-\varepsilon ^{2}\sin ^{2}\varphi _{0}}}{ 1+\cos \varphi _{0}}-\frac{\sin \varphi _{em}\sqrt{1-\varepsilon ^{2}\sin ^{2}\varphi _{em}}}{1+\cos \varphi _{em}}\Biggr]\Biggr\}\mathrm{,} \label{solution1} 
\end{eqnarray} with  
\begin{eqnarray} 
B_{1} &=&\frac{-3^{1/4}\left( 1+2\cosh \frac{\Psi }{3}\right) }{3\left( 4\cosh ^{2}\frac{\Psi }{3}-1\right) ^{1/4}}\mathrm{,}  \nonumber \\ B_{2} &=&\left( \frac{4\cosh ^{2}\frac{\Psi }{3}-1}{3}\right) ^{1/4}\mathrm{, }  \label{B} 
\end{eqnarray} and 
\begin{eqnarray} 
\varepsilon ^{2} &=&\frac{1}{2}+\frac{\sqrt{3}\cosh \frac{\Psi }{3}}{2\sqrt{ 4\cosh ^{2}\frac{\Psi }{3}-1}}\mathrm{,}  \nonumber \\ \varphi \! &=&\!\arccos \!\left\{ \!\frac{\left( 1+z\right) \Omega _{d}\left[ \sqrt{12\cosh ^{2}\frac{\Psi }{3}\!-\!3}-1-2\cosh \frac{\Psi }{3}\!\right] -\!3\Omega _{\rho }}{\left( 1+z\right) \Omega _{d}\left[ \sqrt{12\cosh ^{2} \frac{\Psi }{3}\!-\!3}+1+2\cosh \frac{\Psi }{3}\!\right] +\!3\Omega _{\rho }} \!\!\right\} \!.\mathrm{.}  \label{phi0 and phi} \end{eqnarray} Here $\varphi $ runs in the range $0..\pi /2$. In computing for other values of $\varphi $, we can use the following addition rules for the elliptic integrals: 
\begin{eqnarray} 
F\left( m\pi \pm \varphi ,\varepsilon \right) &=&2mK\pm F\left( \varphi ,\varepsilon \right) \ ,  \nonumber \\ E\left( m\pi \pm \varphi ,\varepsilon \right) &=&2mE\pm E\left( \varphi ,\varepsilon \right) \ ,  \label{addition} 
\end{eqnarray} 
where $K$ and $E$ are the complete elliptic integrals of the first and second kind. Eqs. (\ref{Psi}) and (\ref{solution1})-(\ref{phi0 and phi}) represent the analytical expression of the luminosity distance-redshift relation for FLRW branes with Randall-Sundrum fine-tuning. They are given in terms of the well-known elliptic integrals of first and second kind, and the cosmological parameters $\Omega _{\rho }$, $\Omega _{\lambda }$ and $\Omega _{d}$.

\subsection{Limit of no black hole in the bulk}  

In this subsection we consider the case $\Omega _{d}=0$. The derivation follows closely the steps of the previous subsection, however the formulae are simpler. The auxiliary expression (\ref{Psi}) for $\Psi $ is well defined only for $\Omega _{d}\neq 0$ and we have to address the question how to obtain suitable limits of the results derived for $\Omega _{d}\neq 0.$ For any $\Omega _{d}\ll $ 
\begin{equation} 
\cosh \Psi \approx \frac{27\Omega _{\lambda }\Omega _{\rho }^{2}}{2\Omega _{d}^{3}}\gg 1\mathrm{{\ }.} 
\end{equation} 
But  
\begin{equation} 
\cosh \Psi =\cosh \frac{\Psi }{3}\left( 4\cosh ^{2}\frac{\Psi }{3}-3\right) \approx 4\mathrm{{\ }\cosh ^{3}\frac{\Psi }{3}{\ },} 
\end{equation} 
as $\cosh \left( \Psi /3\right) \gg 1$ also holds. Thus 
\begin{equation} 
\cosh \frac{\Psi }{3}\approx \pm \sinh \frac{\Psi }{3}\approx \frac{3\Omega _{\lambda }^{1/3}\Omega _{\rho }^{2/3}}{2\Omega _{d}}\mathrm{{\ }.} \label{limitPsi} 
\end{equation} 
By employing Eq. (\ref{limitPsi}) in the generic expressions derived in the preceding subsection, we obtain the luminosity distance-redshift relation in a very similar form to Eq. (\ref{solution1}), but with different coefficients:  
\begin{eqnarray} 
d_{L}^{\lambda }\! &=&\!\frac{2\sqrt[4]{3}\left( 1+z\right) \Omega _{\lambda }^{1/6}}{H_{0}\Omega _{\rho }^{2/3}}\Biggl\{\!{}{}{}\!\frac{1}{2}\left( 1- \frac{\sqrt{3}}{3}\right) \left[ F\left( \varphi _{0},\varepsilon \right) -F\left( \varphi _{em},\varepsilon \right) \right]  \nonumber \\ &&+E\left( \varphi _{em},\varepsilon \right) \!-\!E\left( \varphi _{0},\varepsilon \right) \!  \nonumber \\ &&+\!\frac{\sin \varphi _{0}\sqrt{1-\varepsilon ^{2}\sin ^{2}\varphi _{0}}}{ 1+\cos \varphi _{0}}\!-\!\frac{\sin \varphi _{em}\sqrt{1-\varepsilon ^{2}\sin ^{2}\varphi _{em}}}{1+\cos \varphi _{em}}\Biggr\}\!\mathrm{{\ \negthinspace },}  \label{solution2} 
\end{eqnarray} 
where 
\begin{eqnarray} \varepsilon ^{2} &=&\frac{1}{2}+\frac{\sqrt{3}}{4}\mathrm{{\ },}  \nonumber \\ \varphi &=&\arccos \frac{\left( \sqrt{3}-1\right) \Omega _{\lambda }^{1/3}\left( 1+z\right) -\Omega _{\rho }^{1/3}}{\left( \sqrt{3}+1\right) \Omega _{\lambda }^{1/3}\left( 1+z\right) +\Omega _{\rho }^{1/3}}\mathrm{{\ } .}  \label{phi} 
\end{eqnarray} 
Again, $\varphi $ for this case emerges in the limit $\Omega _{d}\rightarrow 0$ from the generic expression Eq. (\ref{phi0 and phi}), by employing Eq. ( \ref{limitPsi}) as in the limiting process expressions of the type $\infty \times 0$ appear.

\subsection{Late-time universe limit}  

In the late-time universe $\rho \ll \lambda $ and in consequence $\Omega _{\lambda }=0$ can be safely assumed. We keep however the dark radiation in the model. Eq. (\ref{chi2}) simplifies considerably, and a straightforward integration gives the luminosity distance - redshift relation  
\begin{equation} 
d_{L}^{d}\left( z\right) =\frac{2\sqrt{1+z} }{H_{0}\Omega _{\rho }} \left( \sqrt{\left( \Omega _{\rho }+\Omega _{d}\right)\left(1+z\right)}-\sqrt{\Omega _{\rho }+\Omega _{d}\left( 1+z\right) }\right) \mathrm{{\ }.}  \label{solution3} 
\end{equation}  

We can also prove that this result emerges as the $\Omega _{\lambda }\rightarrow 0$ limit from the generic results, Eqs. (\ref{Psi}) and (\ref {solution1})-(\ref{phi0 and phi}). When $\Omega _{\lambda }\rightarrow 0$ Eq. (\ref{Psi}) gives $\Psi \rightarrow 0$. Then Eqs. (\ref{phi0 and phi}) and (\ref{B}) give 
\begin{eqnarray} \varepsilon ^{2} &=&1  \nonumber \\ \varphi &=&\arccos \left\{ \frac{-\Omega _{\rho }}{2\left( 1+z\right) \Omega _{d}+\Omega _{\rho }}\right\} \mathrm{{\ },} \\ B_{2} &=&1=-B_{1}\mathrm{{\ }.} 
\end{eqnarray} 
By noting that $E\left( \varphi ,1\right) =\sin \varphi $, we obtain from Eq. (\ref{solution1}):  
\begin{equation} 
d_{L}^{d}=\frac{2\left( 1+z\right) \Omega _{d}^{1/2}}{H_{0}\Omega _{\rho }} \left[ \frac{\sin \varphi _{0}}{1+\cos \varphi _{0}}-\frac{\sin \varphi _{em}}{1+\cos \varphi _{em}}\right] \mathrm{~.} 
\end{equation} 
By inserting the values $\varphi _{em}=\varphi \left( z\right) $ and $ \varphi _{0}=\varphi \left( 0\right) $, we recover the luminosity distance - redshift relation (\ref{solution3}).

\subsection{General relativistic (Einstein-de Sitter) limit}  

The general relativistic limit of the luminosity distance - redshift relation for dust matter and $k=0=\Lambda $ (Einstein-de Sitter model) can be obtained by direct integration of Eq. (\ref{chi2}):  
\begin{equation} 
d_{L}^{GR}\left( z\right) =\frac{2\sqrt{1+z}}{H_{0}\Omega _{\rho }^{1/2}} \left[ \sqrt{1+z}-1\right] \mathrm{{\ },}  \label{solution4} \end{equation} 
It is straightforward to check that the above result stems out from Eq. (\ref {solution3}) by simply switching off the dark radiation.  

The general relativistic limit of the luminosity distance - redshift relation should also emerge in the limit $\Omega _{\lambda }\rightarrow 0$ of Eq. (\ref{solution2}). To see this, we note that when $\Omega _{\lambda }\rightarrow 0$, both $\varphi \rightarrow \pi $ and $\varphi _{0}\rightarrow \pi $. Therefore the elliptic integrals of the first and second kind both tend to finite values, thus the differences evaluated at $ \varphi $ and $\varphi _{0}$ vanish. Then the only terms which should be carefully investigated are the last two terms of Eq. (\ref{solution2}), which are of the type $0/0$. By employing Eq. (\ref{phi}), for the last term we obtain: 
\begin{equation} 
\lim_{\Omega _{\lambda }\rightarrow 0}\Omega _{\lambda }^{1/6}\frac{\sin \varphi _{em}\sqrt{1-\varepsilon ^{2}\sin ^{2}\varphi _{em}}}{1+\cos \varphi _{em}}=\frac{\Omega _{\rho }^{1/6}}{3^{1/4}\sqrt{1+z}}\mathrm{{\ }.} 
\end{equation} 
Accordingly, the second to last term gives  
\begin{equation} 
\lim_{\Omega _{\lambda }\rightarrow 0}\Omega _{\lambda }^{1/6}\frac{\sin \varphi _{0}\sqrt{1-\varepsilon ^{2}\sin ^{2}\varphi _{0}}}{1+\cos \varphi _{0}}=\frac{\Omega _{\rho }^{1/6}}{3^{1/4}}\mathrm{{\ }.} 
\end{equation} 
Adding everything together, we recover the general relativistic result (\ref {solution4}).

\section{Branes with $\Lambda $}  

In this section we discuss certain cases of Randall-Sundrum type brane-worlds with cosmological constant, for which analytical expressions for the luminosity-redshift relation can be found.

\subsection{A brane with analytically integrable luminosity distance-redshift relation}  

If we do not impose the Randall-Sundrum fine-tuning in Eq. (\ref{finetune}) and we keep the brane cosmological constant $\Lambda $, the polynomial in the denominator of the integrand in Eq. (\ref{chi2}) can be simplified for certain values of the dimensionless $\Omega $-s. In particular, if we choose  
\begin{equation} 
\Omega _{d}=0\mathrm{{\ \ \ \ \ \ and \ \ \ \ \ \ }4\Omega _{\lambda }\Omega _{\Lambda }=\Omega _{\rho }^{2}{~,}}  \label{cond} 
\end{equation} the expression under the square root of denominator becomes a quadratic expression, and the integral can be given in terms of elementary functions  \cite{NKfikut}: 
\begin{eqnarray} 
d_{L}^{\Lambda =\kappa ^{2}\lambda /2} &=&\frac{2^{1/3}\left( 1+z\right) }{ 6H_{0}\Omega _{\rho }^{1/3}\Omega _{\Lambda }^{1/6}}\Biggl\{\ln \frac{\left( 1-h+h^{2}\right) \left[ 1+h\left( 1+z\right) \right] ^{2}}{\left( 1+h\right) ^{2}\left[ 1-h\left( 1+z\right) -h^{2}\left( 1+z\right) ^{2}\right] }  \nonumber \\ &&+2\sqrt{3}\arctan \frac{2-h}{\sqrt{3}h}-2\sqrt{3}\arctan \frac{2\left( 1+z\right) -h}{\sqrt{3}h}\Biggr\}~,  \label{solution5} 
\end{eqnarray} 
with $h=\left( \Omega _{\rho }/2\Omega _{\Lambda }\right) ^{1/3}$.  

The first condition (\ref{cond}) merely simplifies the bulk to an anti de Sitter space-time. The second condition (\ref{cond}) by contrast, yields to a much more serious constraint:  
\begin{equation} 
\kappa ^{2}\lambda =2\Lambda  \label{constr} 
\end{equation}  

The second condition (\ref{cond}), together with the constraint (\ref{sumom} ) leads to a quadratic equation for $\Omega _{\lambda }$. For $\Omega _{\rho }=0.27$ this has two solutions \cite{NKfikut}:  
\begin{equation} 
\Omega _{\Lambda }=0.704,\ \ \ \ \ \ \ \ \ \ \Omega _{\lambda }=0.02\,\allowbreak 6  \label{Boti1} 
\end{equation} corresponding to the brane tension\footnote{ All values of the brane tension given in this subsection are in units $ c=1=\hbar $.} $\lambda _{1}=38.375\times 10^{-60}$TeV$^{4}$ and  
\begin{equation} 
\Omega _{\Lambda }=0.026,\ \ \ \ \ \ \ \ \ \ \Omega _{\lambda }=0.704~. \label{Boti2} 
\end{equation} corresponding to the brane tension $\lambda _{2}=1.4173\times 10^{-60}$TeV$ ^{4}$.  

It is interesting to note that while solution (\ref{Boti2}) is ruled out by the recent supernova data, solution (\ref{Boti1}) is quite close to the present observational value of $\Omega _{\Lambda }$ \cite{Liddle}. From a brane point of view, however the value of the brane tension in the model (\ref{Boti1}) is far too small, thus it does not describe our physical world. Indeed, all lower limits set for $\lambda $ are much higher than $\lambda _{2}$.  

In the two-brane model of Ref. \cite{RS1} the minimal brane tension depends on the value of the Planck mass $M_{P}$ and on the characteristic curvature scale of the bulk $l$ as $\lambda _{\min }=3M_{P}^{2}/4\pi l^{2}$ \cite{RS1} . Table-top experiments \cite{tabletop} on possible deviations from Newton's law currently probe gravity at sub-millimeter scales. As a result they constrain the characteristic curvature scale of the bulk to $l\leq 44$ $\mu$m. The brane tension therefore (in units $c=1=\hbar $) is constrained as $ \lambda >715.887$ TeV$^{4}$. (For a detailed discussion see section 6 of \cite {GK}, where a slightly lower bound for the brane tension was derived, based on the previously available estimate $l\leq 0.1$ mm for the characteristic curvature scale of the bulk.) 
Big Bang Nucleosynthesis constraints give a much milder lower limit, $\lambda \gtrsim 1$ MeV$^{4}$ \cite{nucleosynthesis} . An astrophysical limit $\lambda >5\,\times 10^{8}$ MeV$^{4}$ (depending on the equation of state of a neutron star) has also been derived \cite{GM}. This latter value of $\lambda _{\min }$ is in between the two previous lower limits.  

The interpretation of the model (\ref{Boti1}) is the following. The condition (\ref{constr}) on the models with small brane tension implies $\widetilde\Lambda = 0$, thus the bulk becomes flat. 
As such, it has no effect on dynamics and the fifth dimension becomes superfluous. In fact what we face here is a GR model with stiff fluid scaling as $a^{-6}$. 

\subsection{Branes with $\Omega _{d}\ll 1$ and $\Omega _{\protect\lambda  }\ll 1$}  

In this subsection we assume that both $\Omega _{\lambda }$ and $\Omega _{d}$ are small, however we allow for arbitrary values of $\Omega _{\Lambda }$. These assumptions are motivated by observational evidence that at present our universe is extremely close to a $\Lambda $CDM model. A Taylor series expansion of Eq. (\ref{chi2}) gives, to leading order in the small parameters:  
\begin{equation} 
d_{L}^{\Lambda \lambda d}=d_{L}^{\Lambda \mathrm{CDM}}+\Omega _{\lambda }I_{\lambda }+\Omega _{d}I_{d}~,  \label{solution7} 
\end{equation} 
with 
\begin{eqnarray} d_{L}^{\Lambda \mathrm{CDM}} &=&\frac{a_{0}\left( 1+z\right) }{H_{0}} \int_{a_{em}}^{a_{0}}\frac{da}{a^{1/2}\left[ \Omega _{\Lambda }a^{3}+\Omega _{\rho }a_{0}^{3}\right] ^{1/2}}~,  \nonumber \\ I_{\lambda } &=&-\frac{a_{0}^{7}\left( 1+z\right) }{2H_{0}} \int_{a_{em}}^{a_{0}}\frac{da}{a^{7/2}\left[ \Omega _{\Lambda }a^{3}+\Omega _{\rho }a_{0}^{3}\right] ^{3/2}}\mathrm{{\ },}  \nonumber \\ I_{d}^{\left( \alpha \right) } &=&-\frac{a_{0}^{5-\alpha }\left( 1+z\right)  }{2H_{0}}\int_{a_{em}}^{a_{0}}\frac{a^{\alpha -3/2}da}{\left[ \Omega _{\Lambda }a^{3}+\Omega _{\rho }a_{0}^{3}\right] ^{3/2}}\ . 
\end{eqnarray} 
The first expression is the general relativistic luminosity distance - redshift relation in the presence of a cosmological constant (in the $ \Lambda $CDM model). The next two integrals represent the correction functions scaling the small coefficients $\Omega _{\lambda }$ and $\Omega _{d}$.  

All integrands have the same expression $\Omega _{\Lambda }a^{3}+\Omega _{\rho }a_{0}^{3}$ in the denominator. The roots of this cubic polynomial are:  
\begin{eqnarray} 
\alpha &=&-a_{0}\left( \frac{\Omega _{\rho }}{\Omega _{\Lambda }}\right) ^{1/3}\mathrm{{\ },}  \nonumber \\ \beta &=&\frac{a_{0}}{2}\left( \frac{\Omega _{\rho }}{\Omega _{\Lambda }} \right) ^{1/3}\left( 1+i\sqrt{3}\right) \label{roots1} 
\end{eqnarray}
and $\beta ^{\ast }$. 
Then $d_{L}^{\Lambda \mathrm{CDM}}$ can be rewritten as:  
\begin{equation} 
d_{L}^{\Lambda \mathrm{CDM}}=\frac{a_{0}\left( 1+z\right) }{H_{0}\Omega _{\Lambda }^{1/2}}\int_{a}^{a_{0}}\frac{da}{a^{1/2}\left( a-\alpha \right) ^{1/2}\left( a-\beta \right) ^{1/2}\left( a-\beta ^{\ast }\right) ^{1/2}}~. 
\end{equation} The integration can be carried out by employing Eq. (260.00) of \cite{Byrd} and we obtain the result:  
\begin{equation} 
d_{L}^{\Lambda \mathrm{CDM}}\left( z\right) =\frac{\left( 1+z\right) \left[ F\left( \varphi _{0},\varepsilon \right) -F\left( \varphi ,\varepsilon \right) \right] }{3^{1/4}H_{0}\Omega _{\rho }^{1/3}\Omega _{\Lambda }^{1/6}} \mathrm{{\ },}  \label{solution6} 
\end{equation} 
with the variable $\varphi $ and argument $\varepsilon $ of the elliptic integral of the first kind $F\left( \varphi ,\varepsilon \right) $ given by 
\begin{eqnarray} 
\varepsilon ^{2} &=&\frac{1}{2}+\frac{\sqrt{3}}{4}\mathrm{{\ },}  \nonumber \\ \varphi &=&\arccos \frac{\left( 1-\sqrt{3}\right) \Omega _{\Lambda }^{1/3}+\left( 1+z\right) \Omega _{\rho }^{1/3}}{\left( 1+\sqrt{3}\right) \Omega _{\Lambda }^{1/3}+\left( 1+z\right) \Omega _{\rho }^{1/3}}\mathrm{{\ } .}  \label{epphiLCDM} 
\end{eqnarray} 
(Note that $\varepsilon ^{2}$ is the same as in the case $\Omega _{\Lambda }=0=\Omega _{d}$, while $\varphi $ is different. Here $0\leq \varphi \leq \pi /2$ while for other values of $\varphi $, we use Eqs. (\ref{addition}).)  

It is relatively easy to integrate the contribution of the term linear in $ \Omega _{\lambda }$ in terms of the variable $t=a^{3/4}$. After a partial integration meant to reduce the powers in the denominator we employ 
\begin{equation} 
\int_{a}^{a_{0}}\frac{da}{a^{1/2}\left[ a^{3}+\frac{\Omega _{\rho }}{\Omega _{\Lambda }}a_{0}^{3}\right] ^{1/2}}=\frac{\Omega _{\Lambda }^{1/3}}{ 3^{1/4}a_{0}\Omega _{\rho }^{1/3}}\left[ F\left( \varphi _{0},\varepsilon \right) -F\left( \varphi ,\varepsilon \right) \right] \mathrm{{\ },} 
\end{equation} 
and obtain 
\begin{eqnarray} 
I_{\lambda } &=&\frac{1+z}{15H_{0}\Omega _{\rho }^{2}}\left\{ \frac{8\Omega _{\Lambda }+3\Omega _{\rho }}{\left( \Omega _{\Lambda }+\Omega _{\rho }\right) ^{1/2}}-\left( 1+z\right) \frac{8\Omega _{\Lambda }+3\Omega _{\rho }\left( 1+z\right) ^{3}}{\left[ \Omega _{\Lambda }+\Omega _{\rho }\left( 1+z\right) ^{3}\right] ^{1/2}}\right\}  \nonumber \\ &&+\frac{8\Omega _{\Lambda }^{5/6}\left( 1+z\right) }{15\sqrt[4]{3} H_{0}\Omega _{\rho }^{7/3}}\left[ F\left( \varphi _{0},\varepsilon \right) -F\left( \varphi ,\varepsilon \right) \right] \mathrm{\ ,}  \label{IL} \end{eqnarray} 
with the variable $\varphi $ and argument $\varepsilon $ given in Eq. (\ref {epphiLCDM}).  

The last term of Eq. (\ref{solution7}) is much more complicated to evaluate. For $\alpha =1\,\ $\ and $4\,\ $the$\ $source term $\Omega _{d}\ $merely contribute to $\Omega _{\rho }$ and $\Omega _{\Lambda }$, respectively. The more interesting cases are for $\alpha =0\,,~2,~3$. The last term of Eq. (55) for $\alpha =2$ consits of elementary functions: 
\begin{equation} 
I_{d}^{\left( 2\right) }=\frac{1+z}{3\Omega _{\rho }\sqrt{\Omega _{\Lambda }+\Omega _{\rho }\left( 1+z\right) ^{3}}}-\frac{1+z}{3\Omega _{\rho }\sqrt{ \Omega _{\Lambda }+\Omega _{\rho }}}\ , 
\end{equation} 
while $I_{d}^{\left( 0\right) }$ and $I_{d}^{\left( 3\right) }$ are more complicated to evalute, and we give details of the derivation in the Appendix. By passing to the variable $z$ instead of $a$, we obtain: \begin{eqnarray} 
I_{d}^{\left( 0\right) } &=&\frac{\left( 1+z\right) }{3H_{0}\Omega _{\rho }^{2}}\left[ \frac{4\Omega _{\Lambda }+3\Omega _{\rho }}{\sqrt{\Omega _{\Lambda }+\Omega _{\rho }}}-\frac{4\Omega _{\Lambda }+3\left( 1+z\right) ^{3}\Omega _{\rho }}{\left( 1+z\right) \sqrt{\Omega _{\Lambda }+\left( 1+z\right) ^{3}\Omega _{\rho }}}\right]  \nonumber \\ &&-\frac{8\left( 1+z\right) \Omega _{\Lambda }^{1/6}}{3H_{0}\Omega _{\rho }^{5/3}}I_{d}\left( \varphi ,\varepsilon \right) \ . 
\end{eqnarray} 
and 
\begin{eqnarray} 
I_{d}^{\left( 3\right) } &=&-\frac{\left( 1+z\right) }{3H_{0}\Omega _{\rho }} \left[ \frac{1}{\sqrt{\Omega _{\Lambda }+\Omega _{\rho }}}-\frac{1}{\left( 1+z\right) \sqrt{\Omega _{\Lambda }+\left( 1+z\right) ^{3}\Omega _{\rho }}} \right]  \nonumber \\ &&+\frac{2\left( 1+z\right) }{3H_{0}\Omega _{\Lambda }^{5/6}\Omega _{\rho }^{2/3}}I_{d}\left( \varphi ,\varepsilon \right) \ . 
\end{eqnarray} where 
\begin{eqnarray} 
I_{d}\left( \varphi ,\varepsilon \right) &=&\frac{\left( 2+\sqrt{3}\right) }{ \sqrt[4]{3}\left( 1+\sqrt{3}\right) ^{3}}\bigl\{F\left( \varphi _{0},\varepsilon \right) -F\left( \varphi ,\varepsilon \right)  \nonumber \\ &&-\left( 3+\sqrt{3}\right) \left[ E\left( \varphi _{0},\varepsilon \right) -E\left( \varphi ,\varepsilon \right) \right] +\left( 2+\sqrt{3}\right) \left( 3+\sqrt{3}\right)  \nonumber \\ &&\times \left[ \frac{\sin \varphi _{0}\sqrt{1-\varepsilon ^{2}\sin ^{2}\varphi _{0}}}{1+\left( 2+\sqrt{3}\right) \cos \varphi _{0}}-\frac{\sin \varphi \sqrt{1-\varepsilon ^{2}\sin ^{2}\varphi }}{1+\left( 2+\sqrt{3} \right) \cos \varphi }\right] \bigr\}\ .  \label{Idphi} 
\end{eqnarray} 
Thus, the analytic expression of the generic luminosity distance - redshift relation on branes with cosmological constant and small values of $\Omega _{\lambda }$ and $\Omega _{d}$ is given to first order accuracy in these small parameters by Eqs. (\ref{solution7}), (\ref{solution6}), (\ref{IL})-(\ref{Idphi}). 
 
\section{Concluding remarks}  

The main purpose of this paper was to present the analytical formulation of the luminosity distance - redshift relation in the generalized Randall-Sundrum type II brane-world models containing a Weyl fluid either in the form of dark radiation or as radiation leaving the brane and feeding the bulk black holes. We have given the luminosity distance in terms of elementary functions and elliptical integrals of first and second type and we have also shown how the different limits arise from the generic result. Our results hold for:  

(a) Models with Randall-Sundrum fine-tuning ($\Lambda =0$), with or without dark radiation from the bulk and with or without considerable contribution from the energy-momentum squared source terms, discussed in section 4.  

(b) The models discussed in subsection 5.1, obeying $\Lambda =\kappa ^{2}\lambda /2$, integrable in terms of elementary functions and  

(c) Models with a brane cosmological constant, discussed to first order accuracy in both the Weyl fluid and energy-momentum squared sources.  

This last class of models, presented in subsection 5.2 in the latest times of the cosmological evolution are only slightly different from the $\Lambda $ CDM model, as they have $\Omega _{d}\ll 1$ and $\Omega _{\lambda }\ll 1.$ The derived modifications in the luminosity distance - redshift formula then represent corrections to the corresponding formula of the $\Lambda $CDM model.  

While the focus of the present paper is the integrability of the luminosity distance - redshift relation in various brane-world models, in a forthcoming paper \cite{BraneLuminosityDistance2} we will discuss how well the presently available supernova data support the brane-world models with a small amount of Weyl fluid.  

\ack  This work was supported by OTKA grants no. T046939, 69036 and T042509. L \'{A}G and GyMSz were further supported by the J\'{a}nos Bolyai Grant of the Hungarian Academy of Sciences and GyMSz by the Magyary Zolt\'{a}n Higher Educational Public Foundation.  

\appendix  

\section{The evaluation of the integral $I_{d}$}  

We can integrate the last term of Eq. (55) for $\alpha =0,3$, as follows. First we pass to the variable $t=a^{3/2}$ and we perform a partial integration in order to reduce the powers in the denominator of the integrand 
\begin{eqnarray} 
I_{d}^{\left( 0\right) } &=&-\frac{\left( 1+z\right) a_{0}^{5}}{3H_{0}\Omega _{\Lambda }^{3/2}}\int_{a^{3/2}}^{a_{0}^{3/2}}\frac{dt}{t^{4/3}\left[ t^{2}+ \frac{\Omega _{\rho }a_{0}^{3}}{\Omega _{\Lambda }}\right] ^{3/2}}  \nonumber \\ &=&\frac{\left( 1+z\right) \Omega _{\Lambda }^{1/2}}{3H_{0}a_{0}\Omega _{\rho }^{2}}\left[ \frac{4t^{2}+3\frac{\Omega _{\rho }a_{0}^{3}}{\Omega _{\Lambda }}}{t^{1/3}\sqrt{t^{2}+\frac{\Omega _{\rho }a_{0}^{3}}{\Omega _{\Lambda }}}}\right] _{a^{3/2}}^{a_{0}^{3/2}}  \nonumber \\ &&-\frac{8\left( 1+z\right) \Omega _{\Lambda }^{1/2}}{9H_{0}a_{0}\Omega _{\rho }^{2}}\int_{a^{3/2}}^{a_{0}^{3/2}}\frac{t^{2/3}dt}{\sqrt{t^{2}+\frac{ \Omega _{\rho }a_{0}^{3}}{\Omega _{\Lambda }}}}\ , 
\end{eqnarray} 
and 
\begin{eqnarray} 
I_{d}^{\left( 3\right) } &=&-\frac{\left( 1+z\right) a_{0}^{2}}{3H_{0}\Omega _{\Lambda }^{3/2}}\int_{a^{3/2}}^{a_{0}^{3/2}}\frac{t^{2/3}dt}{\left[ t^{2}+ \frac{\Omega _{\rho }a_{0}^{3}}{\Omega _{\Lambda }}\right] ^{3/2}}  \nonumber \\ &=&-\frac{\left( 1+z\right) }{3H_{0}a_{0}\Omega _{\Lambda }^{1/2}\Omega _{\rho }}\left[ \frac{t^{2}}{t^{1/3}\sqrt{t^{2}+\frac{\Omega _{\rho }a_{0}^{3}}{\Omega _{\Lambda }}}}\right] _{a^{3/2}}^{a_{0}^{3/2}}  \nonumber \\ &&+\frac{2\left( 1+z\right) }{9H_{0}a_{0}\Omega _{\Lambda }^{1/2}\Omega _{\rho }}\int_{a^{3/2}}^{a_{0}^{3/2}}\frac{t^{2/3}dt}{\sqrt{t^{2}+\frac{ \Omega _{\rho }a_{0}^{3}}{\Omega _{\Lambda }}}}\ . 
\end{eqnarray} 
By a change of the integration variable to $x=\Omega _{\Lambda }^{1/3}t^{3/2}/\Omega _{\rho }^{1/3}a_{0}$ we can employ Eq. (260.52) of  \cite{Byrd} in order to evaluate the remaining integral:  
\begin{eqnarray} 
\int_{a^{3/2}}^{a_{0}^{3/2}}\frac{t^{2/3}dt}{\sqrt{t^{2}+\frac{\Omega _{\rho }a_{0}^{3}}{\Omega _{\Lambda }}}} &=&\frac{3a_{0}\Omega _{\rho }^{1/3}}{ 2\Omega _{\Lambda }^{1/3}}\int_{\frac{\Omega _{\Lambda }^{1/3}a}{\Omega _{\rho }^{1/3}a_{0}}}^{\frac{\Omega _{\Lambda }^{1/3}}{\Omega _{\rho }^{1/3}} }\frac{x^{2}dx}{\sqrt{x^{3}+1}}  \nonumber \\ &=&\frac{3a_{0}\Omega _{\rho }^{1/3}}{\Omega _{\Lambda }^{1/3}}I_{d}\left( \varphi ,\varepsilon \right) \ . 
\end{eqnarray} 
Here $I_{d}\left( \varphi ,\varepsilon \right) $ is given by Eqs. (\ref {Idphi}).

\end{document}